\documentclass{desyproc}

\usepackage{amssymb}

\newcommand{\gagamma}{g_{a\gamma}}
\newcommand{\ckcs}{counts~keV$^{-1}$~cm$^{-2}$~s$^{-1}$}

\newcommand{\zgz}{$^1$}
\newcommand{\usc}{$^2$}
\newcommand{\tri}{$^3$}
\newcommand{\lbnl}{$^4$}
\newcommand{\cern}{$^5$}
\newcommand{\demok}{$^6$}
\newcommand{\irfu}{$^7$}
\newcommand{\tud}{$^8$}
\newcommand{\rbi}{$^9$}
\newcommand{\llnl}{$^{10}$}
\newcommand{\desy}{$^{11}$}
\newcommand{\mpi}{$^{12}$}
\newcommand{\bgu}{$^{13}$}
\newcommand{\inr}{$^{14}$}
\newcommand{\nps}{$^{15}$}
\newcommand{\pat}{$^{16}$}

\begin{document}
%------------------------------------
\title{The International Axion Observatory (IAXO)}

%for single authors the superscripts are optional
%\author{{\slshape Axel Lindner$^1$, Konstantin Zioutas$^{2,3}$}\\[1ex]
%$^1$Deutsches Elektronen-Synchrotron (DESY), Hamburg, Germany\\
%$^2$Physics Department, University of Patras, Patras, Greece\\
%$^3$European Organization for Nuclear Research (CERN), Geneve, Switzerland}

\author{\slshape  I.~G.~Irastorza\zgz, F.~T.~Avignone\usc, G.~Cantatore\tri, S.~Caspi\lbnl, J.~M.~Carmona\zgz, T.~Dafni\zgz, M.~Davenport\cern, A.~Dudarev\cern, G.~Fanourakis\demok, E.~Ferrer-Ribas\irfu, J.~Gal\'an\zgz$^,$\irfu, J.~A.~Garc\'ia\zgz,
T.~Geralis\demok, I.~Giomataris\irfu, S.~Gninenko\inr, H.~G\'omez\zgz,
D.~H.~H.~Hoffmann\tud, F.~J.~Iguaz\irfu, K.~Jakov\v{c}i\'{c}\rbi,
   M.~Kr\v cmar\rbi, B.~Laki{\' c}\rbi, G.~Luz\'on\zgz, A.~Lindner\desy, M.~Pivovaroff\llnl, T.~Papaevangelou\irfu,  G.~Raffelt\mpi, J.~Redondo\mpi, A.~Rodr\'iguez\zgz, S.~Russenschuck\cern, J.~Ruz\cern, I.~Shilon\cern$^,$\bgu,
   H.~Ten~Kate\cern, A.~Tom\'as\zgz, S.~Troitsky\inr,  K.~van~Bibber\nps, J.~A.~Villar\zgz, J.~Vogel\llnl, L.~Walckiers\cern, K.~Zioutas\pat \\[1ex]
\zgz Laboratorio de F\'{\i}sica Nuclear y Astropart\'{\i}culas, Universidad
de Zaragoza, Zaragoza, Spain\\
\usc Department of Physics and Astronomy, University of South Carolina, Columbia, SC, USA\\
\tri Instituto Nazionale di Fisica Nucleare (INFN), Sezione di Trieste and Universit\'a di Trieste, Trieste, Italy\\
\lbnl Lawrence Berkeley National Laboratory, Berkeley, CA 94720, USA \\
\cern CERN, Geneva, Switzerland\\
\demok National Center for Scientific Research "Demokritos", Athens, Greece \\
\irfu IRFU, Centre d'\'Etudes Nucl\'eaires de Saclay (CEA-Saclay), Gif-
sur-Yvette, France\\
\tud Technische Universit\"at Darmstadt, IKP, Darmstadt, Germany\\
\rbi Rudjer Bo\v{s}kovi\'{c} Institute, Zagreb, Croatia\\
\llnl Lawrence Livermore National Laboratory, Livermore, CA, USA\\
\desy DESY, Hamburg, Germany\\
\mpi Max-Planck-Institut f\"ur Physik, Munich, Germany \\
\bgu Physics Department, Ben-Gurion University of the Negev, Beer-Sheva 84105, Israel \\
\inr Institute for Nuclear Research (INR), Russian Academy of Sciences, Moscow, Russia\\
\nps Naval Postgraduate School, Monterey, CA, USA \\
\pat University of Patras, Patras, Greece}

% if the proceedings are available online (e.g. at Indico)
% please enter the contribution ID or file_name below for the DOI
%\contribID{32}
\contribID{Irastorza_Igor}

% TO THE CONFERENCE EDITORS:
% please update the following information
% before sending the template to the authors
% \confID{800}  % if the conference is on Indico uncomment this line
\desyproc{DESY-PROC-2011-04}
\acronym{Patras 2011} % if you want the Acronym in the page footer uncomment this line
\doi  % if there is an online version we will register DOIs

\maketitle

\begin{abstract}
The International Axion Observatory (IAXO) is a new generation axion helioscope aiming at a sensitivity to the axion-photon coupling of
$\gagamma\gtrsim {\rm few} \times 10^{-12}$~GeV$^{-1}$, i.e. 1--1.5 orders of magnitude beyond the one currently achieved by CAST. The project relies on improvements in magnetic field volume together with extensive use of x-ray focusing optics and low background detectors, innovations already successfully tested in CAST. Additional physics cases of IAXO could include the detection of electron-coupled axions invoked to explain the white dwarf cooling, relic axions, and a large variety of more generic axion-like particles (ALPs) and other novel excitations at the low-energy frontier of elementary particle physics. This contribution is a summary of our recent paper \cite{Irastorza:2011gs}.
\end{abstract}

%%%%%%%%%%%%%%%%%%%%%%%%%%%%%%%%%%%%%%%%%%%%%%%%%%%%%%%%%%%%%%%%%%%%%%
\section{Introduction}                        \label{sec:introduction}
%%%%%%%%%%%%%%%%%%%%%%%%%%%%%%%%%%%%%%%%%%%%%%%%%%%%%%%%%%%%%%%%%%%%%%

The Peccei-Quinn (PQ) mechanism of dynamical symmetry
restoration~\cite{Peccei:1977ur,Peccei:1977hh} stands out as the most
compelling solution of the strong CP problem. Central to the PQ mechanism
is the axion~\cite{Weinberg:1977ma,Wilczek:1977pj},
the Nambu-Goldstone boson of a new spontaneously broken symmetry
U(1)$_{\rm PQ}$. The properties of axions
allow them to be produced in the early universe as coherent field
oscillations and as such to provide all or part of the cold dark
matter~\cite{Sikivie:2006ni,Wantz:2009it}.

It is still possible to find these ``invisible axions'' in
realistic search experiments and in this way test a fundamental
aspect of QCD. The generic $a\gamma\gamma$ vertex allows for
axion-photon conversion in external electric or magnetic fields in
analogy to the Primakoff effect for neutral pions. As shown in
1983 by Pierre Sikivie, the smallness of the axion mass allows
this conversion to take place coherently over macroscopic
distances, compensating for the smallness of the interaction
strength~\cite{Sikivie:1983ip}. Especially promising is to use the
Sun as a source for axions produced in its interior by the
Primakoff effect. Directing a strong dipole magnet toward the Sun
allows one to search for keV-range x-rays produced by axion-photon
conversion, a process best visualized as a particle oscillation
phenomenon~\cite{Raffelt:1987im} in analogy to neutrino flavor
oscillations. Three such helioscopes have been built, in
Brookhaven~\cite{Lazarus:1992ry}, Tokyo~\cite{Moriyama:1998kd} and
at CERN~\cite{Zioutas:1998cc}. The CERN Axion Solar Telescope
(CAST) has just finished a 8-year long data taking period,
having strongly improved on previous experiments and even
surpassed astrophysical limits in some range of parameters,
although axions have not been found.

We have shown \cite{Irastorza:2011gs} that large improvements in magnetic field volume,
x-ray focusing optics and detector backgrounds with respect to CAST
are possible. Based on these improvements, and on the experience gathered within CAST, we propose the International Axion Observatory (IAXO), a new generation axion
helioscope. IAXO could search for axions that are 1--1.5 orders of
magnitude more weakly interacting that those allowed by current CAST
constraints. It appears conceivable to surpass the SN~1987A
constraint on the axion mass, $m_a\lesssim10$--20~meV, test the white-dwarf (WD) cooling hypothesis~\cite{Isern:2010wz}, and 
explore a substantial part of uncharted axion territory experimentally. Moreover, IAXO would explore other more generic models of weakly interacting sub-eV particles (WISPs) \cite{Jaeckel:2010ni}, in particular some ALPs models that have been invoked in the context of several unexplained astrophysical observations. Equipped with microwave cavities, this setup could also aim at detecting relic axions \cite{Baker:2011na}.

%If the ambitious goals defined in our study can be
%achieved, a much larger range of realistic axion models can be
%probed and it is even conceivable that one can reach a sensitivity
%corresponding to $m_a$ in the 10~meV range. This mass range would be
%significant in several ways. The energy-loss limit from SN~1987A
%suggests that QCD axions have $f_a\gtrsim10^9$~GeV or
%$m_a\lesssim10$--20~meV as mentioned earlier. Moreover, if axions also
%interact with electrons, axions nearly saturating the SN~1987A limit
%could explain the apparent anomalous energy loss of white
%dwarfs~\cite{Isern:2010wz}. On
%the experimental side, if the magnet length is 10~m as in CAST, the
%sensitivity loss caused by the axion-photon momentum transfer begins
%at $m_a\gtrsim20$~meV. In other words, if one can ``cross the axion
%line'' at around this mass means that one can probe a large range of
%axion models without buffer gas filling or with only few simple
%pressure settings.

\section{Experimental setup and expected sensitivity}

\begin{figure}[t] \centering
\includegraphics[width=4cm]{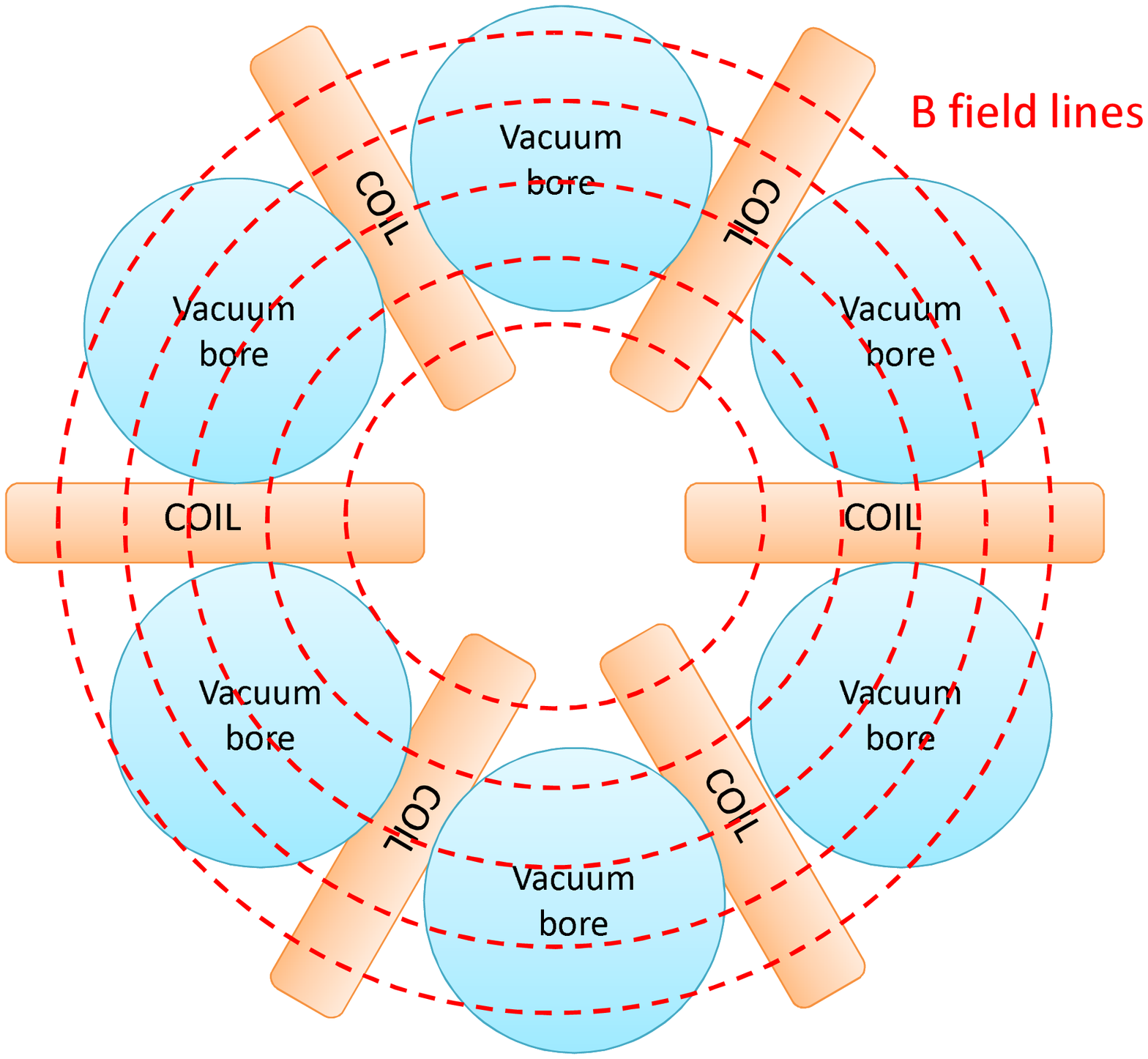}
\includegraphics[width=8cm]{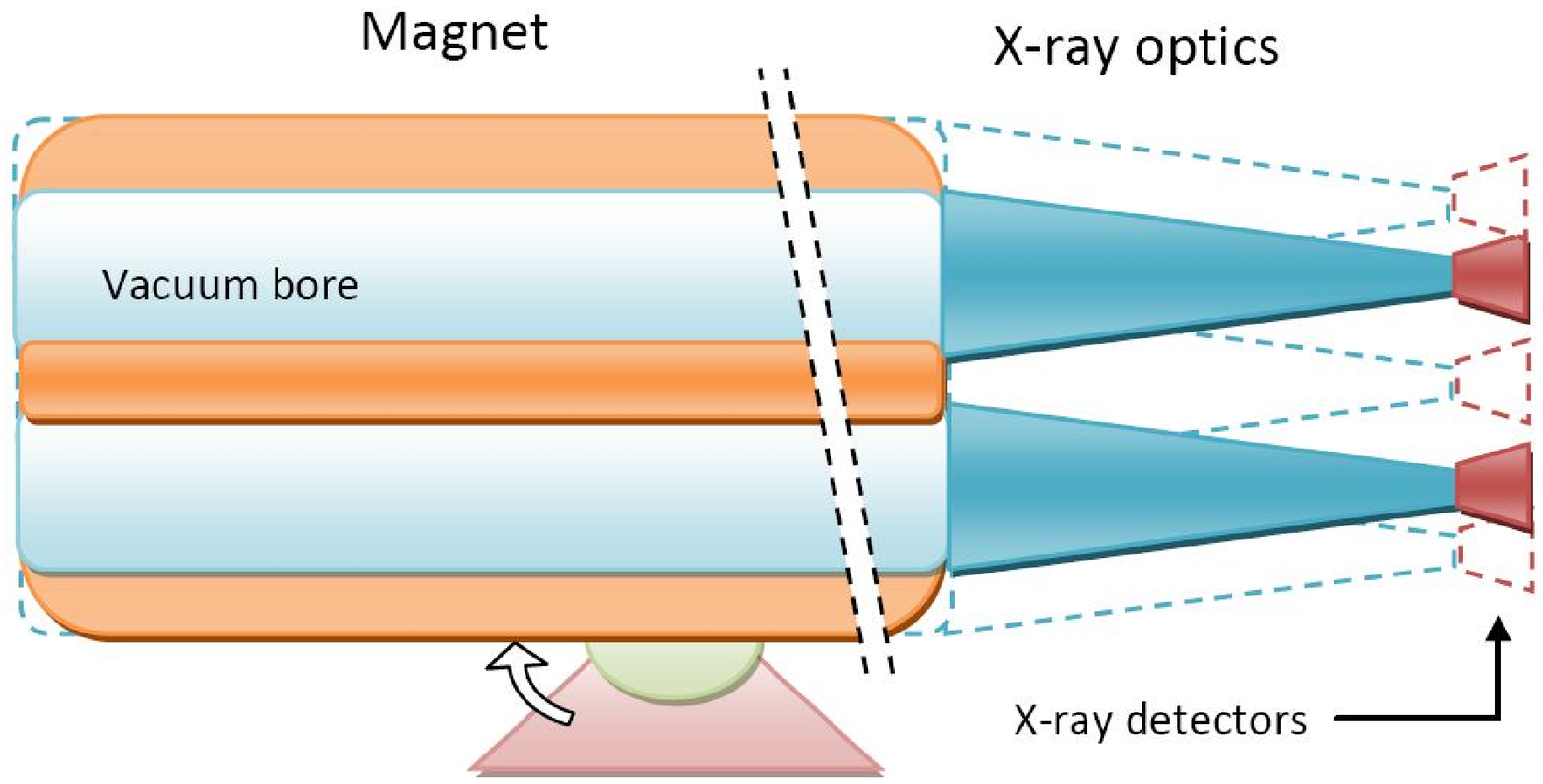}\hspace{2pc}%
\caption{\label{fig:sketch_NGAH} Possible conceptual arrangement
for IAXO. On the left we show the cross section of the IAXO
toroidal magnet, in this example with six coils and bores. On the
right the longitudinal section with the magnet, the optics
attached to each magnet bore and the x-ray detectors.}
\end{figure}

IAXO will follow the basic conceptual layout of an enhanced axion helioscope seen in figure~\ref{fig:sketch_NGAH}, which shows a toroidal design for the magnet, together with X-ray optics and detectors attached to each of the magnet bores. The improvements anticipated for each of the experimental parameters of the helioscope were quantified in \cite{Irastorza:2011gs}, organized in four scenarios (IAXO 1 to 4) ranging from most conservative to most optimistic values (see table 1 of \cite{Irastorza:2011gs}). These values are justified by several considerations on the magnet, x-ray optics and detectors, that are briefly outlined in the following, but we refer to \cite{Irastorza:2011gs} for a detailed discussion.

The magnet parameters are the ones contributing mostly to the helioscope's figure of merit. The CAST success has relied, to a large extent, on the
availability of the first class LHC test magnet which was recycled to become part of the CAST helioscope.
While going beyond CAST magnet's $B$ or $L$ is difficult, the improvement may come however in the cross section area, which
in the case of the CAST magnet is only $3\times10^{-3}$~m$^2$. Substantially larger cross sections can be achieved, although one
needs a different magnet configuration. It is an essential part of
our proposal that a new magnet must be designed and built
specifically for this application, if one aims at a substantial
step forward in sensitivity. A toroidal configuration for the IAXO magnet is being studied with a total cross section area 
$A$ of up to few m$^2$, while keeping the product of
$BL$ close to levels achieved for CAST.

Another area for improvement will be the x-ray optics. Although CAST has
proven the concept, only one of the four CAST magnet bores is
equipped with optics. The use of focusing power in the entire magnet
cross section $A$ is implicit in the figures of merit defined in \cite{Irastorza:2011gs}, and
therefore the improvement obtained by enlarging $A$ comes in part
because a correspondingly large optic is coupled to the magnet.
Here the challenge is not so much achieving exquisite focusing or near-unity
reflectivity but the availability of cost-effective x-ray
optics of the required size. IAXO's optics specifications can be met by a dedicated fabrication effort based on segmented glass substrate optics like the ones of HEFT or NuSTAR \cite{nustar}.

Finally, CAST has enjoyed the sustained
development of its detectors towards lower backgrounds during its
lifetime. The latest generation of Micromegas detectors in CAST are
achieving backgrounds of $\sim$~$5\times10^{-6}$~\ckcs.
This value is already a factor 20 better than the backgrounds recorded during the
first data-taking periods of CAST. Prospects for reducing this
level to $10^{-7}$ \ckcs\ or even lower appear feasible.

\begin{figure}[t] \centering
\includegraphics[height=6.5cm]{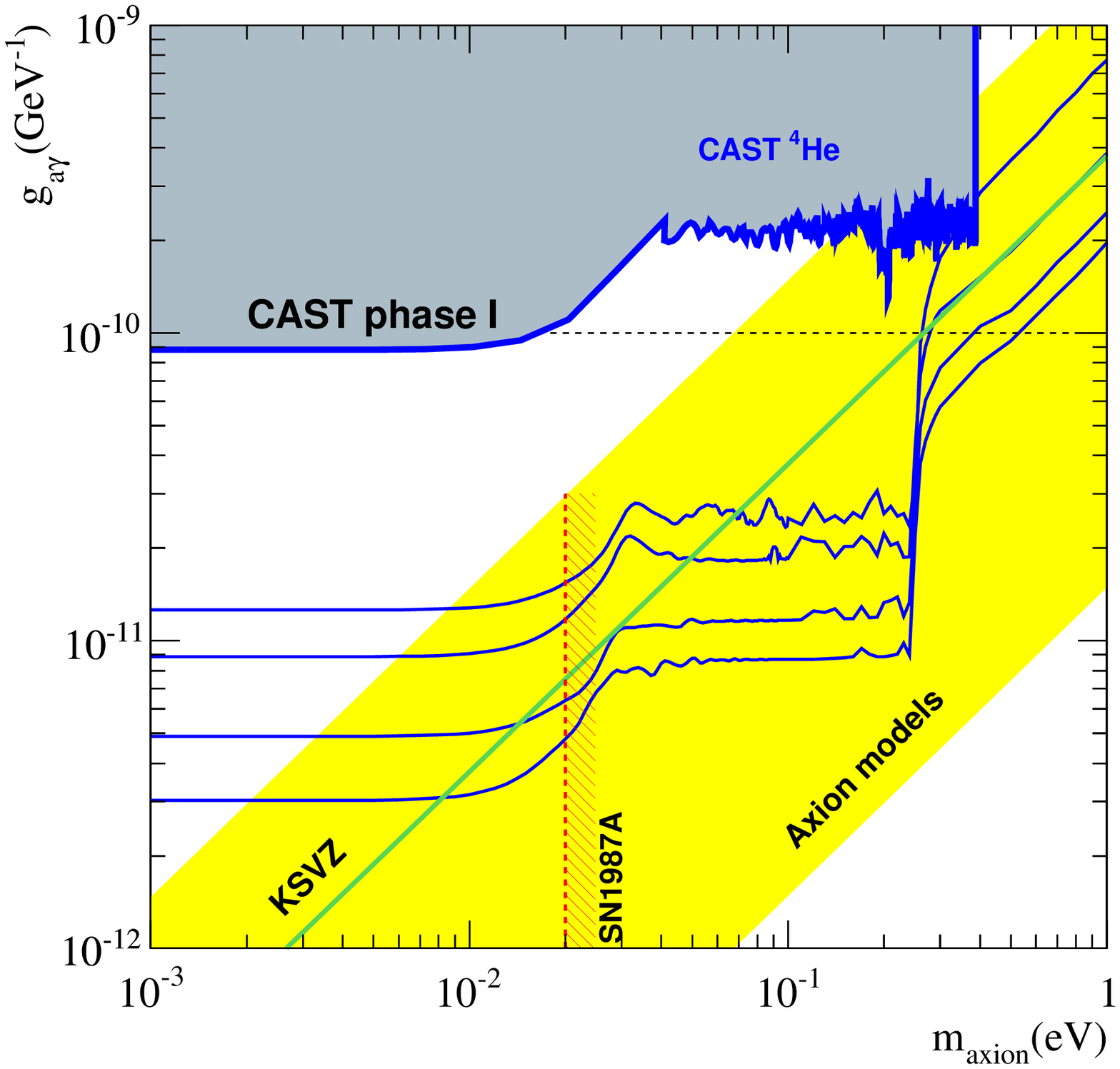}
\includegraphics[height=6.5cm]{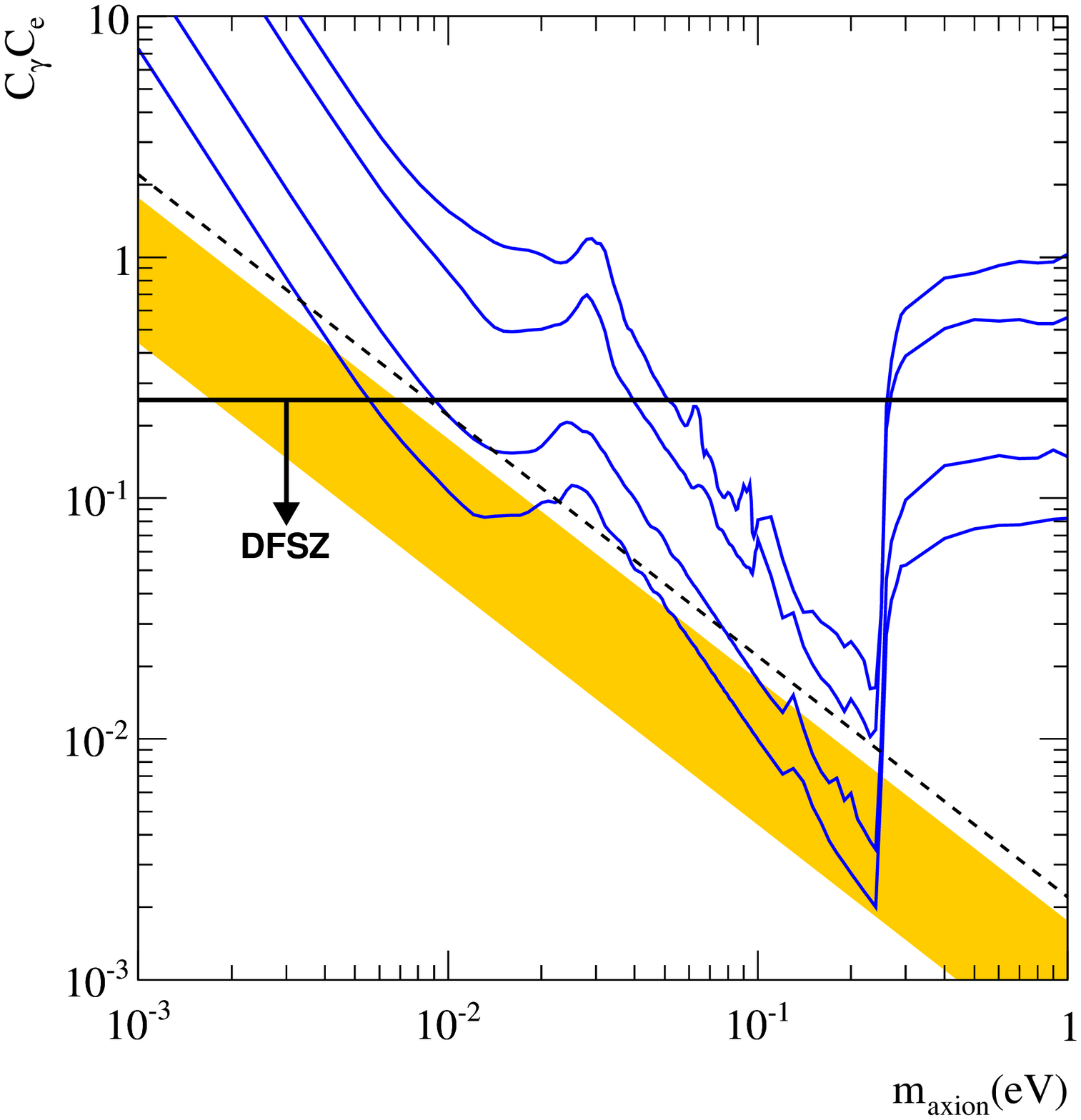}\hspace{2pc}%
\caption{\label{fig:scenarios} LEFT: The parameter space for hadronic axions and ALPs. The CAST limit, some other limits, and the range of PQ models (yellow band) are also shown. The blue lines indicate the sensitivity of the four scenarios discussed in the text% and table \ref{scenarios}.
. RIGHT: The expected sensitivity regions of the same four scenarios in the parameter space of non-hadronic axions with both electron and photon coupling. The orange band represents the region motivated by WD cooling, and the dashed line along the diagonal the red giants bound on the electron coupling. See \cite{Irastorza:2011gs} for details.}
% by red gianIn GUT
%models $C_\gamma$ is fixed to $0.75$ and we show the bound on the
%electron coupling ($C_e$) from red giants (dashed line along the diagonal) and the
%region motivated by WD cooling (orange band).  DFSZ models lie
%below the horizontal line $C_\gamma C_e< 0.25$.}
\end{figure}

The computed sensitivities of each of the four IAXO scenarios are represented by the family of blue lines in figure~\ref{fig:scenarios}, both for hadronic axions (left) and non-hadronic ones (right). They include two data taking campaigns for each of the scenarios: one three years long performed without buffer gas (analogous to CAST I), and another three years long period with varying amounts of $^4$He gas inside the magnet bore (analogous to CAST II, although without the need to use $^3$He). 
%This second phase is responsible for the step in the sensitivity line from mass of $\sim$0.05 eV up to 0.25 eV. This range is given by the gas density %range chosen for this calculation of 0 to 1 bar of $^4$He at room temperature. Of course, a shorter density range could be chosen, thus allowing for a %mass scan correspondingly shorter but more sensitive in $\gagamma$.
In general, IAXO sensitivity lines go well beyond current CAST sensitivity for hadronic axions and progressively
penetrate into the decade $10^{-11}$--$10^{-12}$ GeV$^{-1}$, with
the best one approaching $10^{-12}$ GeV$^{-1}$. They are sensitive
to realistic QCD axion models at the 10 meV scale and exclude a good
fraction of them above this. For non-hadronic axions, IAXO sensitivity lines penetrate in the DFSZ model region, approaching or even surpassing the red-giant constraints. Most relevantly, the IAXO 3 and IAXO 4 scenarios start probing the region of parameter space highlighted by the cooling of WDs.

%
%
%
%
%
%Figure~\ref{fig:mmback} shows the background levels achieved by
%the CAST Micromegas detectors along the experiment's lifetime.
%Solid dots, representing the nominal levels achieved in CAST data
%taking periods show a decrease in background by a factor 20 since
%the start of the experiment. Last generation of Micromegas, made
%with the microbulk fabrication technique, with radiopure
%components and properly shielded (see
%figure~\ref{fig:mmpictures}), present a background of 5--10$\times 10^{-6}$ \ckcs.
%
%
%
%\begin{figure}[ht!]
%    \centering
%     \includegraphics[width = .70\textwidth, angle = 0]{MMback.eps}
%      \caption{Background levels of Micromegas detectors over the years.
%      Black points represents nominal values in CAST data taking campaigns.
%      Squared red points correspond to data taken in special shielding conditions in the Canfranc underground laboratory.}
%    \label{fig:mmback}
%\end{figure}
%
%
%
%
%\section*{Bibliography}

%\bibliographystyle{h-physrev3}
%\bibliographystyle{iopart-num}
%\bibliographystyle{JHEP}
%\bibliography{../../BibTeX/igorbib,pivovaroff,redondobib}

% ****************************************************************************
% BIBLIOGRAPHY AREA
% ****************************************************************************

\begin{footnotesize}

\end{footnotesize}

% ****************************************************************************
% END OF BIBLIOGRAPHY AREA
% ****************************************************************************

\end{document}